\documentclass[showpacs,preprintnumbers,amsmath,amssymb,20pt]{revtex4}
\usepackage{graphicx}
\usepackage{color}
\usepackage{bm}
\newcommand{\beq}{\begin{equation}}
\newcommand{\eeq}{\end{equation}}
\newcommand{\bqa}{\begin{eqnarray}}
\newcommand{\eqa}{\end{eqnarray}}
\begin{document}
\title{
Formation of vortex and antivortex polarization states in ferroelectric films}
\author{Manas Kumar Roy, Sushanta Dattagupta}
\affiliation{Indian Institute of Science Education and Research - Kolkata,\\
Mohanpur, PIN: 741252, India.\\}
\begin{abstract}
An effort based on the kinetic model allowed by Landau-Ginzburg-Devonshire theory is presented here. The formation of   vortex, antivortex states in ferroelectric films are reported and theoretically explained. Phase transition in toroidal moment is observed. Vortex-antivortex-vortex (V-AV-V) triplet is generated  by applying  inhomogeneous transverse static localized fields in x-y plane. A specific boundary condition is used to keep the spirit of vortex state i.e. the net polarization across the boundary is always zero. 
\end{abstract}
\maketitle
\section{Introduction}
\phantom{xxxx}Circularly ordered structural phase is denoted as vortex state which is  found in ferromagnets \cite{kittel, Shinjo}, superconductors \cite{Glad}, Bose Einstein condensates\cite{Marti}, and in ultracold Fermi gases \cite{Bote}.
Kittel \cite{kittel} has shown the possibility of circular ordering due to boundary conditions of magnetic spin vectors in any ferroic material.  Existence of such vortex orders was extended  into  ferroelectrics by Ginzburg et.al. \cite{Ginz}  in  1984.\\ 
\phantom{xxxx}In 2004, Naumov et al \cite{Nau1} defined a different order parameter 'toroidal moment' ($\vec{\mathbf{G}}$) through  a first principle calculation, where $\vec{\mathbf{G}}$ is the vector sum of  cross products between local dipole moments and corresponding  position vectors of the dipoles. This toroidal moment passes through a phase transition during temperature variation. In lower temperature, polarization  forms a toroidal domain configuration. They show that   FE vortex states are a promisingly smart material that can fasten memory performance and   of higher memory capacity by five orders of magnitude compared to memory of same size. This explanation has accelerated the ab initio explorations in the theoretical domain  of polarization vortex structures. Gruverman et al \cite{Gruv} in 2008 experimentally observed doughnout shape ferroelectric domain pattern and further numerically analyzed it by a model based on Landau- Lifshitz-Ginzburg equations. Existence of such vortex polar states depends on the size as well as shape of the nanodot, where strain and depolarizing field play crucial role\cite{Nau2}. Our recent observation by time dependent Landau Ginzburg theory has demnonstrated the formation of  boundary condition induced vortex polarization states \cite{Manas}. An intrinsic coupling condition  between two transverse in plane components can reorient a head to tail 90$^\circ$ to a circularly symmetric vortex domain. \\
\phantom{xxxx} But, magnetic, BE, superconducting systems, antivortex is also mentioned. Recently Roy et al \cite{Roy} observed vortex-antivortex chains in permalloy rings. The difference between ferroelectric and ferromagnetic vortex dynamics, nearest neighbour and depolrizing field is very strong in the former system compared to the latter. But vortex and antivortex pair can not form as ferroelectric toroidal moment does not break time reversal symmetry \cite{Prosa}. Whereas in magnetic troidal moment $T=1/{2N}\sum_i^N{r_i \times m_i}$, (here $m_i$ are the local magnetic dipoles at sited in N lattices) breaks both the time and space symmetry \cite{Prosa,Gorba}. A recent  ab-initio study by Prosandev et al \cite{Prosa} has  shown that external inhomogeneous field can control the toroidal moment. Here we report the results of a study of the vortex antivortex vortex (VAV) switching kinetics in the ultrathin ferroelectric films, which are almost two dimensional. As we are applying in plane electric field, the cross product does not break time as well as spatial symmetry. \\
\phantom{xxxx}In this Letter,  we explore a bilateral excersise between the two in plane polarization components under the tutelage of anisotropic electric fields. External electric fields are applied in the diagonal axis of the 2D square ferrolectric lattice as  shown by Roy et al \cite{Roy}  a chain of the  magnetization distribution around the vortex antivortex cores of the VAV structure in permalloy rings. The temperature is below critical tempertature.  This is qualitatively different from the kinetics of extrinsic switching,  where a conducting atomic force micriscopic tip switches ferrolectric films into a 180$^\circ$ domain. Here dipoles are switched out of plane.\\
\phantom{xxxx}In the framework of Landau Ginzburg Devonshire  phenomenology for a most symmetric case, the spatial-temporal evolution of the polarization components in a 2D  film of the second order ferroelectric is described by these two coupled nonlinear parabbolic equation \cite{Sdork, Ishi}:
\bqa
\frac{\partial P_x}{\partial t}=\Gamma \left[(AP_x-BP_x^3-CP_xP_y^2)+ \delta\nabla^2P_x\right]+E_x,\\
\frac{\partial P_y}{\partial t}=\Gamma\left[(AP_y-BP_y^3-CP_yP_x^2)+ \delta\nabla^2P_y\right]+E_y.
\eqa
where A is a pheneomenological temeperature term defined for second order transition as $A=A_0(T_c-T)$. Below the critical temperature $T_c$, ferroelectric phase will appear. $A_0$, B, C are positive constantsis. $E_x$ and $E_y$ are   components of effective fields appearing from external electric fields and depolrizing fields  acting in x and y direction respectively.Here $\Gamma$ is phenomenological kinetic coefficient. Corresponding toroidal moment $\vec{G}$ is defined as:
\begin{equation}
\vec{G}=\frac{1}{2N}\sum_i^N \vec{R}_i\times \vec{P}_i
\end{equation}
For a 2D film, no z component of polarization is present. Therefore, $\vec{G}$ only have z component and that will be $G_z=\frac{1}{2N}\sum_i^N (x_iP_{y_i}-y_iP_{x_i})$\\
We adopt  with the following  boundary conditions:\\
\begin{table}[h]
\begin{tabular}{ | c | c | c | c |c | }
\hline
Polarization states & $x=-L$ & $x=L$ & $y=-L$ & $y=L$ \\
\hline
Vortex & $P_x$=0,$P_y$=-1  & $P_x$=0,$P_y$=1 & $P_x$=1,$P_y$=0 & $P_x$=-1,$P_y$=0 \\
\hline
Anti-vortex & $P_x$=0,$P_y$=1  & $P_x$=0,$P_y$=-1 & $P_x$=1,$P_y$=0 & $P_x$=-1,$P_y$=0 \\
\hline
\end{tabular}
\caption{Boundary conditions considered}
\end{table} 
\\
We have probed the electric field at two specific sites $S_1$ and $S_2$  along $x=y$ axis in the following way:\\
\begin{table}[h]
\begin{tabular}{ | c | c |  c |c | }
\hline
 & $S_1$ & $S_2$  \\
\hline
$E_x$  & -1 & 1    \\
\hline
$E_y$  & 1  & -1  \\
\hline
\end{tabular}
\caption{External electric field}
\end{table}
\\
\phantom{xxxx}In order to realize phase transition in toroid moment $G_z$, we simulate Eq.1 and Eq.2 with the boundary condition specified in Table 1 for vortex state. We have taken total 100 sites i.e. N=100. Figure 1(a) shows net toroid moment $G_z$ versas a temperature dependent parameter $A^\prime=A_0(T-Tc)$ ($A^\prime=-A$). This result is similar in spirit of the work by Naumov et al who has already submitted the evidence of phase transition in $G_z$. It is evident from figure 1(a), that at temperature $T<T_c$ ($A^\prime<0$), we observe a nonzero toroidal moment, whereas for for $T>T_c$ i.e.$A^\prime>0$, $G_z$ will sharply be zero. In a time domain of finite difference, we simulated Eq.1 and Eq.2. We initiate  $A^\prime=-2$, and formed the vortex after 800 time steps. Then we gradually increased the value of $A^\prime$ by 0.1. This process continued until $A^\prime=2$. Figure 1(b) and 1(c) shows the equilibrium vortex and antivortex states for $A^\prime=-1$ respectively. \\
\phantom{xxxx}We are now planning to construct VAV triplet. Earlier, a combined effect of a critical depolarizing field and an open circuit electrical boundary condition has been mentioned, this combination  can convert a out-of-plane polarized nanodot into a  nonzero toroidal moment. Figure 2 shows a schematic  external field probing diagram. Fields are in the plane i.e. only have two components along x and y plane. No z component is present. That is also true for the depolarizing field along z axis. We have taken a system of square ferroelectric lattice in a form of 2D film with 160 points.  External fields are applied at two places S1 and S2. S1 and S2 are in diagonal axis (along line x=y), placed equal distance from center. Coordinates of S1 and S2 are \{15,15\} and \{25,25\} respectively. We used the same boundary condition used for creating the vortex state as mentioned in table 1. Figure 3 (a), 3(b),  3(c), and 3(d) are showing formation of VAV triplet for time steps 1, 50, 100, and 400 respectively. If the fields are withdrawn simultaneously, this triplet structure will converge very slowly to a single vortex. System will again capture the rotational symmetry which was earlier disturbed by the anisotropic electric fields.\\
\phantom{xxxx}Fig.4(a) and 4(b) show equilibrium $P_x$, and $P_y$ profile in xy plane. Figure.4(c) describes the spatial distribution of toroidal moment$G_z$ in xy plane. It is clear in Fig.4(c), there is a chain of vortex-antivortex-vortex has been formed. Figure 4(d) is the plot of $P_x$ and $P_y$ along the x=y diagonal line. These apparently periodic lines cut at three points predicting VAV locations. Both the $P_x$, and $P_y$ attain zero polarization values at these points. It reminds us the  superposition of  two periodic waves with same frequency and amplitude but  $\pi$ phase difference.\\
\phantom{xxxx}In summary, this Letter presents a method  of formation of VAV triplet in thin ferroelectric films. Simple Landau Divonshire Ginzburg theory in a  kinetic form, and special boundary conditions can produce vortex antivortex states. Phase transition in toroid moment has been observed. We proposed, if external fields of anisotropic nature are skillfully probed in  small ferroelectric thin films, system, VAV triplet can be formed. Application of two electric place in close vicinity can produce three distinctly written zones.  This can be utilized during memory application, and used as a source for further development of FeRAM.\\
\phantom{xxxx}The authors acknowledge  Dr.Shamik Sarkar and Dr. Jaita Paul for fruitful discussions.  Authors also thank the unknown reviewer of their previous work \cite{Manas} for his constructive criticism.

\newpage
\begin{figure}[h]
\begin{center}
\includegraphics[width=1\textwidth]{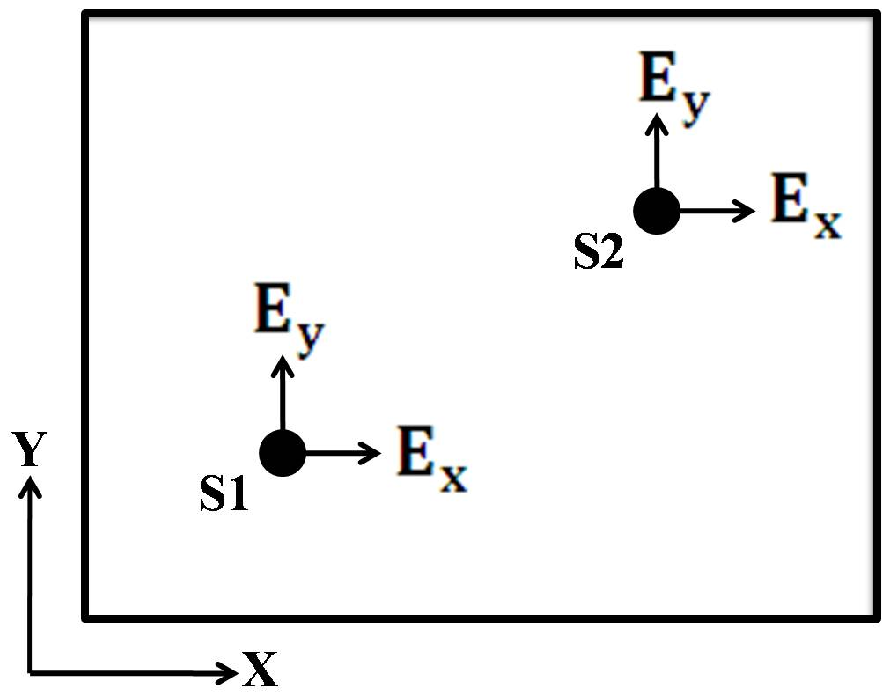}
\caption{\label{fig:1} Schematic diagram shows how and where the electric field can be applied.}
\end{center}
\end{figure}
\begin{figure}[h]
\begin{center}
\includegraphics[width=1\textwidth]{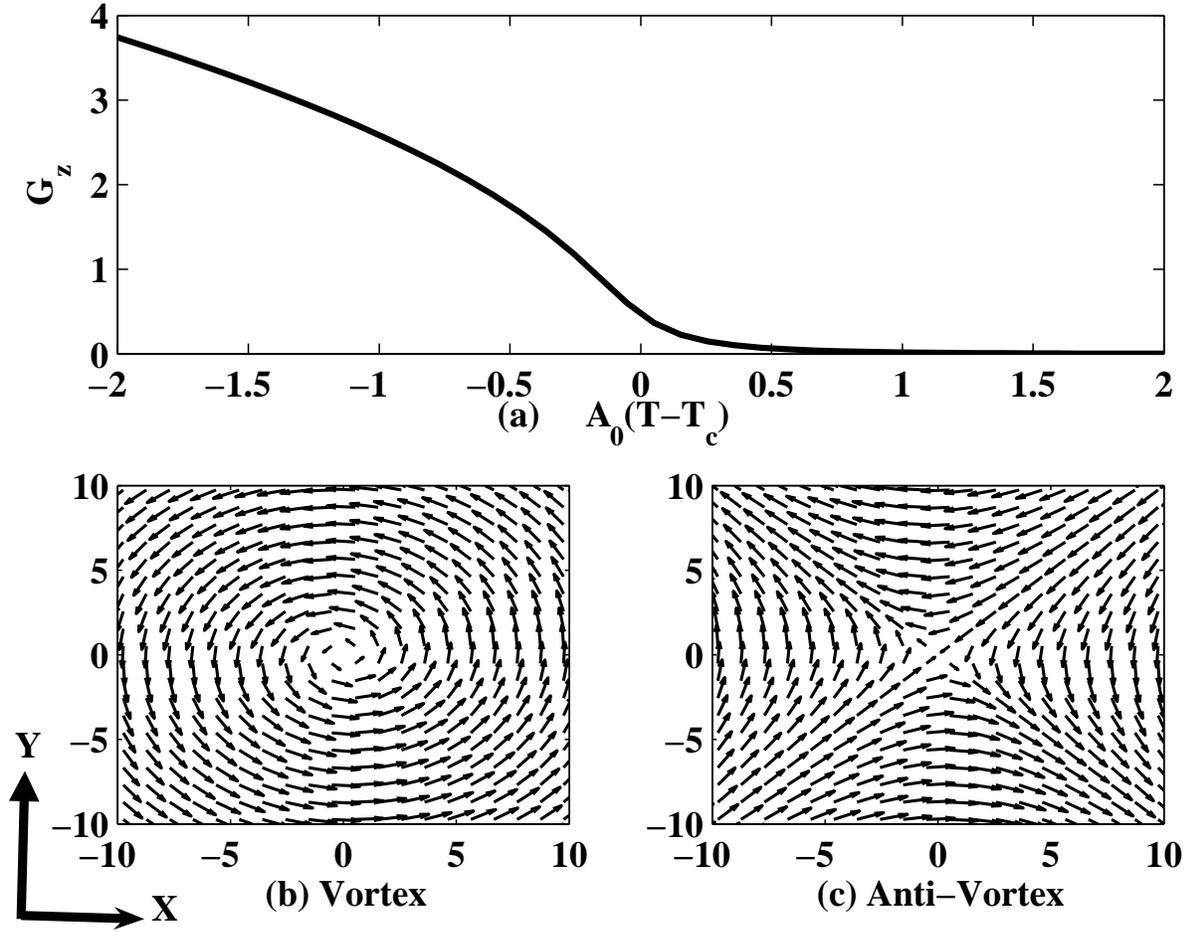}
\caption{\label{fig:2} (a) Phase transition in the z component of toroidal moment $\vec{\mathbf{G}}$ where $\vec{\mathbf{G}}=\frac{1}{2N}\sum_i^N \vec{R}_i\times \vec{P}_i$.  (b) Vortex and (c) antivortex.}
\end{center}
\end{figure}
\begin{figure}[h]
\begin{center}
\includegraphics[width=1\textwidth]{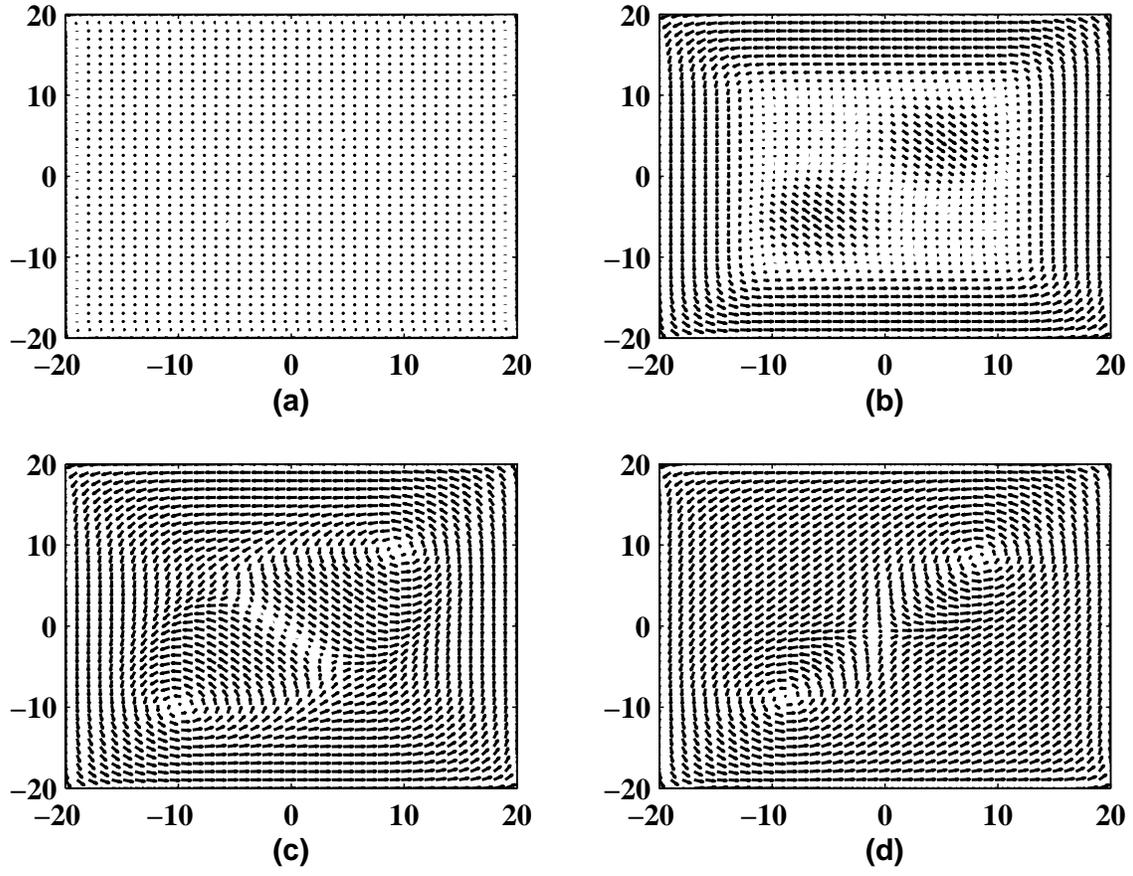}
\caption{\label{fig:3} Vortex antivortex and vortex formation at time steps (a)1, (b)50, (c)100 and (d) 400.}
\end{center}
\end{figure}
\begin{figure}[h]
\begin{center}
\includegraphics[width=1\textwidth]{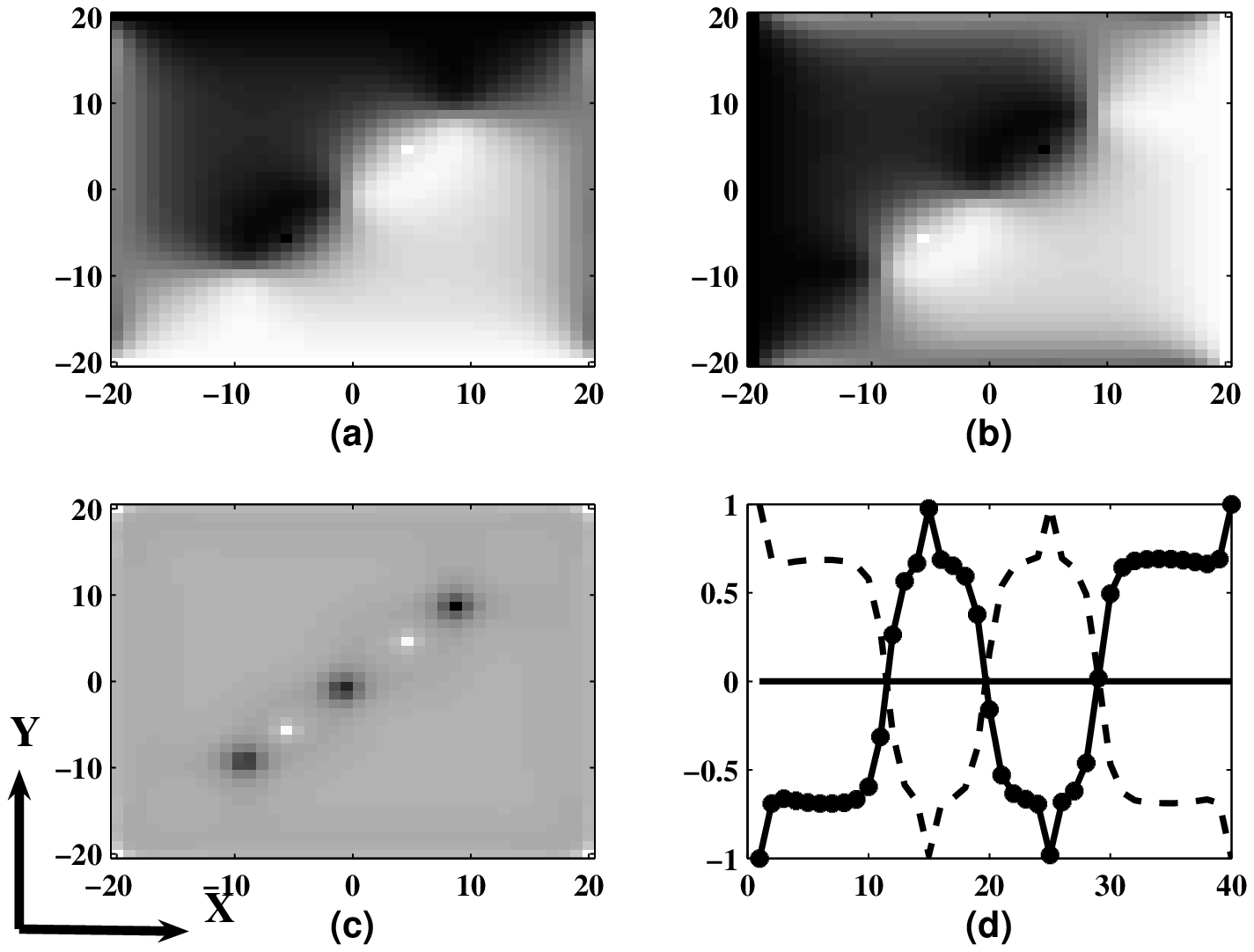}
\caption{\label{fig:4}(Color online) Spatial configuration of (a) $P_x$, (b)$P_y$ and $\vec{\mathbf{G}}$ at equilibrium. (d) shows the spatial profile of $P_x$ and $P_y$ along diagonal and there crossover. }
\end{center}
\end{figure}


\begin{thebibliography}{000}
\bibitem{kittel} C. Kittel, Phys. Rev. 70, 965 (1946).
\bibitem{Shinjo} T. Shinjo, T. Okuno, R. Hassdorf, K. Shigeto, and T. Ono,
Science 289, 930 (2000).
\bibitem{Glad}V. N. Gladilin, J. Tempere, J. T. Devreese, W. Gillijns, and V. V. Moshchalkov, Phys. Rev. B. 80, 054503 (2009).
\bibitem{Marti}J.-P. Martikainen, K.-A. Suominen1, L. Santos, T. Schulte, and A. Sanpera, Phys. Rev. A 64, 063602 (2001). 
\bibitem{Bote}S. S. Botelho and C. A. R. Sá de Melo, Phys. Rev. Lett. 96, 040404 (2006).
\bibitem{Ginz}
Ginzburg, V. L.; Gorbatsevich, A. A.; Kopayev, Y. V.; Volkov, B. A. Solid State Commun. 1984, 50, 339.
\bibitem{Nau1}
 Naumov, I. I.; Bellaiche, L., Fu, H. X. Nature 2004, 432, 737.
\bibitem{Gruv}
Gruverman, A.; Wu, D.; Fan, H.-J.; Vrejoiu, I.; Alexe, M.; Harrison,
R. J.; Scott, J. F. J. Phys.: Condens. Matter 2008, 20, 342201.
\bibitem{Nau2}
Naumov, I.; Bratkovsky, A. M. Phys. ReV. Lett. 2008, 101, 107601.
\bibitem{Manas}M. K. Roy, S. Sarkar, and S. Dattagupta, Appl. Phys. Lett. 95, 192905 (2009).
\bibitem{Gorba}
Gorbatsevich, A. A.; Kopayev, Y. V. Ferroelectrics 1994, 161, 321.
\bibitem{Prosa}S. Prosandeev, I. Ponomareva, I. Kornev, I. Naumov, and L. Bellaiche, Phys. Rev. Lett. 96, 237601 (2006).
\bibitem{Roy}P. E. Roy, J. H. Lee, T. Trypiniotis, D. Anderson, G. A. C. Jones, D. Tse, and C. H. W. Barnes, Phys. Rev. B 79, 060407 (R) (2009).
\bibitem{Sdork}A.S. Sidorkin, \textit{Domain Structure in Ferroelectrics and Related Materials}, (Cambridge International Science Publishing 2006).
\bibitem{Ishi}Y. Ishibashi and E. Salje, Journal of the Physical Society of Japan Vol. \textbf{71}, No. 11, November, 2002, pp. 2800-2803.
\end{thebibliography}
\end{document}